\documentclass[aps,twocolumn,showpacs,superscriptaddress]{revtex4}

\usepackage{graphicx}
\usepackage{bm}
\usepackage{amsmath}
\usepackage{dcolumn}%

\newcommand{\eg}{\ensuremath{e_{g}}}
\newcommand{\tg}{\ensuremath{t_{2g}}}
\newcommand{\dxy}{\ensuremath{d_{xy}}}
\newcommand{\dxz}{\ensuremath{d_{xz}}}
\newcommand{\dyz}{\ensuremath{d_{yz}}}
\newcommand{\dzz}{\ensuremath{d_{3z^2-1}}}
\newcommand{\dxxyy}{\ensuremath{d_{x^2-y^2}}}

\newcommand{\mb}{\ensuremath{\mu_{\text{B}}}}

\newcolumntype{/}{D{/}{/}{2,2}}  
\newcolumntype{.}{D{.}{.}{0}}  

\begin{document}

\title{Resonant inelastic x-ray scattering in layered trimer iridate
  Ba$_4$Ir$_3$O$_{10}$: the density functional approach }

\author{D.A. Kukusta}

\affiliation{G. V. Kurdyumov Institute for Metal Physics of the
  N.A.S. of Ukraine, 36 Academician Vernadsky Boulevard, UA-03142
  Kyiv, Ukraine}

\author{L.V. Bekenov}

\affiliation{G. V. Kurdyumov Institute for Metal Physics of the
  N.A.S. of Ukraine, 36 Academician Vernadsky Boulevard, UA-03142
  Kyiv, Ukraine}

\author{V.N. Antonov}

\affiliation{G. V. Kurdyumov Institute for Metal Physics of the
  N.A.S. of Ukraine, 36 Academician Vernadsky Boulevard, UA-03142
  Kyiv, Ukraine}

\affiliation{Max-Planck-Institute for Solid State Research,
  Heisenbergstrasse 1, 70569 Stuttgart, Germany}

\date{\today}

\begin{abstract}

We have investigated the electronic structure of Ba$_4$Ir$_3$O$_{10}$ within
the density-functional theory (DFT) using the generalized gradient
approximation while considering strong Coulomb correlations (GGA+$U$) in the
framework of the fully relativistic spin-polarized Dirac linear muffin-tin
orbital band-structure method. Ba$_4$Ir$_3$O$_{10}$ has a quasi-2D structure
composed of buckled sheets, which constitute corner-connected Ir$_3$O$_{12}$
trimers containing three distorted face-sharing IrO$_6$ octahedra. The Ir
atoms are distributed over two symmetrically inequivalent sites: the center of
the trimer (Ir$_1$) and its two tips (Ir$_2$). The Ir$_1$ $-$ Ir$_2$ distance
within the trimer is quite small and equals to 2.58\AA\, at low
temperature. As a result, the clear formation of bonding and antibonding
states at the Ir$_1$ site occurs. The large bonding-antibonding splitting
stabilizes the $d_{yz}$-orbital-dominant antibonding state of $t_{2g}$ holes
and produces a wide energy gap at the Fermi level. However, the energy gap
opens up only with taking into account strong Coulomb correlations at the
Ir$_2$ site.  Therefore, we have quite a unique situation when the insulating
state is driven by both the dimerization at the Ir$_1$ site and Mott
insulating behavior at the Ir$_2$ one. We have investigated resonant inelastic
x-ray scattering (RIXS) spectra at the Ir $L_3$ edge. The calculated results
are in good agreement with experimental data. The RIXS spectrum possesses
several sharp features below 2.1 eV corresponding to transitions within the Ir
{\tg} levels. The excitation located from 2.1 to 4.6 eV is due to {\tg}
$\rightarrow$ {\eg} and O$_{2p}$ $\rightarrow$ {\tg} transitions. The wide
structure situated at 6.2$-$12 eV appears due to charge transfer and O$_{2p}$
$\rightarrow$ {\eg} transitions. We have also presented comprehensive
theoretical calculations of the RIXS spectrum at the oxygen $K$ edge.

\end{abstract}

\pacs{75.50.Cc, 71.20.Lp, 71.15.Rf}

\maketitle

\section{Introduction}

\label{sec:introd}

Interplay between orbital, spin, charge carriers, and lattice degrees of
freedom has been a fascinating subject for the condensed-matter physics
community for the last few decades. The 5$d$ transition-metal compounds
possess the same order of magnitude of on-site Coulomb repulsion $U$,
spin-orbit coupling (SOC), and the crystal-field energy \cite{WCK+14}. SOC in
such systems splits the {\tg} orbitals into a quartet ($J_{\rm{eff}}$ =
$\frac{3}{2}$) and a doublet ($J_{\rm{eff}}$ = $\frac{1}{2}$)
\cite{JaKh09,CPB10,WCK+14}. In 5$d^5$ (Ir$^{4+}$) iridium oxides, such as
Sr$_2$IrO$_4$, the quartet $J_{\rm{eff}}$ = $\frac{3}{2}$ is fully occupied,
and the relatively narrow $J_{\rm{eff}}$ = $\frac{1}{2}$ doublet, occupied by
one electron, can be splitted by moderate Hubbard $U_{\rm{eff}}$ with opening
a small band gap called the relativistic Mott gap
\cite{KJM+08,MAV+11,AUU18,AKB24}. Iridates have been at the center of
intensive research in recent years for novel phenomena, such as topological
insulators \cite{QZ10,Ando13,WBB14,BLD16}, Mott insulators
\cite{KJM+08,KOK+09,WSY10,MAV+11}, Weyl semimetals \cite{WiKi12,GWJ12,SHJ+15},
and quantum spin liquids (QSLs) \cite{KAV14,Bal10,SaBa17}.

The specific physical properties of iridates strongly depend on local
geometry. The cases most widely discussed in the literature are the geometry
with MO$_6$ octahedra (M is a transition-metal ion) sharing a common oxygen (a
common corner) or two common oxygens (octahedra with a common edge).  The
first case takes place, for example, in the LaMnO$_3$ perovskite or layered
systems as La$_2$CuO$_4$. The case with a common edge occurs in many layered
systems with triangular lattices such as NaCoO$_2$ and LiNiO$_2$. The features
of spin-orbit systems in both these cases were studied in detail (see, e.g.,
Ref. \cite{book:Khomski14}).  The third typical geometry, which can also be
very often met in many real materials, is the case of octahedra with a common
face (three common oxygens). Actually, there are many transition-metal
compounds with this geometry \cite{KKS+15}.  Such materials include, for
example, the series of perovskites Ba$_3$MIr$_2$O$_9$ (M = Y$^{3+}$,
Sc$^{3+}$, In$^{3+}$, Lu$^{3+}$) \cite{SDH06,DoHi04}, which show diverse
ground-state properties. These systems possess blocks of two face-sharing
IrO$_6$ octahedra separated by MO$_6$ octahedra (which have common corners
with IrO$_6$). They have two equivalent Ir sites, therefore, the oxidation
state of Ir in these systems is fractional: Ba$_3$M$^{3+}$Ir$_2^{4.5+}$O$_9$.
This fractional oxidation state of Ir (accompanied by a unique
crystallographic site) can lead to fascinating ground states.  At high
temperatures, where the mixed-valence Ir$_2$O$_9$ dimers can be seen as
isolated, the magnetic susceptibility deviates from the conventional
Curie-Weiss behavior \cite{DMH02,SDH06,NaRa17} suggesting nontrivial
temperature evolution of the local magnetic moment. At low temperatures,
interactions between the dimers become important, and signatures of frustrated
magnetic behavior \cite{ZAS+17} including possible formation of a spin-liquid
ground state have been reported \cite{DMO+17}.

Another example with two blocks of face-sharing octahedra is the
Ba$_5$AlIr$_2$O$_{11}$ barium oxide. The crystal structure of this compound
consists of AlO$_4$ tetrahedra and IrO$_6$ octahedra. The latter octahedra
share a face and develop Ir$_2$O$_9$ dimers. Novel properties are expected
from this structural arrangement in addition to those driven by SOC.  Opposite
to the Ba$_3$MIr$_2$O$_9$ systems, Ba$_5$AlIr$_2$O$_{11}$ features dimer
chains of two inequivalent octahedra occupied by tetravalent Ir$^{4+}$
(5$d^5$) and pentavalent Ir$^{5+}$ (5$d^4$) ions,
respectively. Ba$_5$AlIr$_2$O$_{11}$ is a Mott insulator that undergoes a
subtle structural phase transition near $T_S$ = 210 K and a transition to
magnetic order at $T_M$ = 4.5 K. The ferrimagnetic (FiM) state below $T_M$ is
highly anisotropic and resilient to a strong magnetic field (up to 14 T) but
is susceptible to even modest hydrostatic pressure \cite{TWY+15}.

Recently, unconventional electronic and magnetic ground states have been
reported in compounds with basic units of Ir trimers, i.e., three face-sharing
IrO$_6$ octahedra, which are much less explored so far
\cite{CCG+00,NgCa19,CZZ+20}. Among them, Ba$_4$Ir$_3$O$_{10}$ with a Ir 5$d^5$
nominal atomic configuration and a monoclinic $P2_1/a$ structure
\cite{CHW+21}. It has a quasi-2D structure composed of buckled sheets, which
constitute corner-connected Ir$_3$O$_{12}$ trimers containing three distorted
face-sharing IrO$_6$ octahedra. The Ir atoms are distributed over two
symmetrically inequivalent sites: Ir$_1$, at the center of the trimers, is at
the Wyckoff position 2$a$, and two outer Ir$_2$ atoms are at the Wyckoff
position 4$e$. The shortest Ir-Ir bond is the one between Ir$_1$ and Ir$_2$
within the trimers, which has a length of $\sim$2.58 \AA\,
\cite{CZZ+20,SSL02}, shorter than in the metallic iridium ($\sim$2.71 \AA\,
\cite{book:Wyck63}) and in the dimer perovskites Ba$_3$InIr$_2$O$_9$
($\sim$2.65 \AA\, \cite{DMO+17}) and Ba$_5$AlIr$_2$O$_{11}$ ($\sim$2.73 \AA\,
\cite{MuLa89}).

Here we report a theoretical investigation of the electronic and magnetic
structures of the Ba$_4$Ir$_3$O$_{10}$ perovskite. There are several
experimental investigations of its physical properties with some controversial
results and conclusions
\cite{WM91,KRD+11,SK+12,KKS+15,CZZ+20,CZH+20,CHW+21,SSF+22}.  The transport
and magnetization studies by Cao {\it et al.} \cite{CZZ+20} found no magnetic
order down to 0.2 K despite the Curie-Weiss temperatures up to $-$766 K and an
enormously large averaged frustration parameter ($f$ $\sim$ 2000). In
contrast, the material shows linear behavior in the low-temperature magnetic
heat capacity, resembling a gapless quantum spin liquid (QSL). In fact, the
ground state is rather fragile in regard to external perturbations including
sample growth conditions \cite{CZH+20,CHW+21} and chemical doping: even a
slight substitution of 1 or 2\% isovalent Sr for Ba recovers the magnetic
order and destroys the signatures of the QLS \cite{CZZ+20}. Together, these
reports suggest that the magnetic ground state properties of
Ba$_4$Ir$_3$O$_{10}$ are extremely delicate. Although Gao {\it et al.}
\cite{CZZ+20} has speculated that Ba$_4$Ir$_3$O$_{10}$ could be either a
Luttinger liquid QSL or a 2D QSL such as spinon Fermi surface states, the
authors of Ref. \cite{SSF+22} argue that Ba$_4$Ir$_3$O$_{10}$ possesses a 1D
spinon continuum. Therefore, instead of forming an isotropic QSL state, the
magnetic frustration can effectively reduce the system dimension, suppressing
the magnetic order and realizing deconfined spinons in a unique way.

Cao {\it et al.} \cite{CZH+20} investigated the physical properties of
structurally altering Ba$_4$Ir$_3$O$_{10}$ via application of a magnetic field
during materials growth. The magnetic field itself was remarkably very weak,
no stronger than 0.06 Tesla. Structurally, the field-altered single crystal
exhibits an elongation in the $b$ axis with only slight changes in the $a$ and
$c$ axes as well as the increase of both Ir-Ir bond distances within each
trimer and the Ir-O-Ir bond angle between the trimers. They found that small
changes in the lattice constants and bond angles cause disproportionately
large electronic changes. The quantum liquid in the non-altered
Ba$_4$Ir$_3$O$_{10}$ is replaced with the AFM state in the field-altered
Ba$_4$Ir$_3$O$_{10}$, emphasizing the critical role of SOI and Coulomb
correlations in this class of quantum materials.

Recently, Chen {\it et al.}  \cite{CHW+21} utilized together resonant elastic
x-ray scattering (REXS), magnetometry, and specific heat to investigate the
ground state properties of Ba$_4$Ir$_3$O$_{10}$ single crystals. In contrast
to previous results \cite{CZZ+20}, the authors discovered clear magnetic order
in Ba$_4$Ir$_3$O$_{10}$. The scattering data revealed two consecutive second
order structural and magnetic transitions at $T_S$ $\sim$ 142 K and $T_N$
$\sim$ 25 K, respectively. The two transitions are consistent with the kinks
in the anisotropic magnetization data, which are qualitatively similar to the
data from the previous report \cite{CZZ+20}. Shen {\it et al.}  \cite{SSF+22}
by measuring the RIXS spectra at the Ir $L_3$ edge found clear evidence of
magnon excitations at low energy ($\sim$0.1 eV) in Ba$_4$Ir$_3$O$_{10}$, and
therefore, this oxide couldn't be characterized as QSL.

In this paper, we focus our attention on the RIXS properties of
Ba$_4$Ir$_3$O$_{10}$. The RIXS method has shown remarkable progress as a
spectroscopic technique to record the momentum and energy dependence of
inelastically scattered photons in complex materials. RIXS rapidly became the
forefront of experimental photon science \cite{AVD+11,GHE+24}. It combines
spectroscopy and inelastic scattering to probe the electronic structure of
materials. This method is bulk sensitive, polarization dependent, as well
as element and orbital specific \cite{AVD+11}.  It permits direct measurements
of phonons, plasmons, single-magnons, and orbitons, as well as other many-body
excitations in strongly correlated systems, such as cuprates, nickelates,
osmates, ruthenates, and iridates, with complex low-energy physics and exotic
phenomena in the energy and momentum space.

There is great progress in the RIXS experiments over the past decade. The most
calculations of the RIXS spectra of various materials have been carried out
using the atomic multiplet approach with some adjustable parameters and the
number of theoretical first-principle calculations of RIXS spectra is
extremely limited. In this paper, we report a theoretical investigation from
the first principles of the RIXS spectra of Ba$_4$Ir$_3$O$_{10}$. Recently,
the RIXS measurements have been successfully performed at the Ir $L_3$ edge in
Ba$_4$Ir$_3$O$_{10}$ by Shen {\it et al.} \cite{SSF+22} in the energy range up
to 5 eV. In addition to the elastic peak centered at zero energy loss, the
spectrum consists of several peaks below 2 eV and a peak at 3.8 eV. The
authors also discovered low-energy magnetic excitations at $\sim$0.1 eV. They
also presented the RIXS spectra at the Ir $L_3$ edge of the isovalently doped
(Ba$_{1-x}$Sr$_x$)$_4$Ir$_3$O$_{10}$ ($x$ = 0.02).

We carry out here a detailed study of the electronic structure and RIXS
spectra of Ba$_4$Ir$_3$O$_{10}$ in terms of the density functional theory. Our
study sheds light on the important role of band structure effects and
transition metal 5$d$ $-$ oxygen 2$p$ hybridization in the spectral properties
of 5$d$ oxides. We use the {\it ab initio} approach using the fully
relativistic spin-polarized Dirac linear muffin-tin orbital band-structure
method. Both the generalized gradient approximation (GGA) and the GGA+$U$
approach are used to assess the sensitivity of the RIXS results to different
treatment of the correlated electrons.

The paper is organized as follows. The crystal structure of
Ba$_4$Ir$_3$O$_{10}$ and computational details are presented in
Sec. II. Section III presents the electronic and magnetic structures of
Ba$_4$Ir$_3$O$_{10}$. In Sec. IV, the theoretical investigation of the RIXS
spectrum of Ba$_4$Ir$_3$O$_{10}$ at the Ir $L_3$ and oxygen $K$ edges is
presented, the theoretical results are compared with experimental
measurements. Finally, the results are summarized in Sec. V.

\section{Computational details}
\label{sec:details}

\subsection{X-ray magnetic circular dichroism} 

Within the one-particle approximation, the absorption coefficient
$\mu^{\lambda}_j (\omega)$ for incident x-ray polarization $\lambda$ and
photon energy $\hbar \omega$ can be determined as the probability of
electronic transitions from initial core states with the total angular
momentum $j$ to final unoccupied Bloch states

\begin{eqnarray}
\mu_j^{\lambda} (\omega) &=& \sum_{m_j} \sum_{n \bf k} | \langle \Psi_{n \bf k} |
\Pi _{\lambda} | \Psi_{jm_j} \rangle |^2 \delta (E _{n \bf k} - E_{jm_j} -
\hbar \omega ) \nonumber \\
&&\times \theta (E _{n \bf k} - E_{F} ) \, ,
\label{mu}
\end{eqnarray}
where $\Psi _{jm_j}$ and $E _{jm_j}$ are the wave function and the
energy of a core state with the projection of the total angular
momentum $m_j$; $\Psi_{n\bf k}$ and $E _{n \bf k}$ are the wave
function and the energy of a valence state in the $n$-th band with the
wave vector {\bf k}; $E_{F}$ is the Fermi energy.

$\Pi _{\lambda}$ is the electron-photon interaction
operator in the dipole approximation
\begin{equation}
\Pi _{\lambda} = -e \mbox{\boldmath$\alpha $} \bf {a_{\lambda}},
\label{Pi}
\end{equation}
where $\bf{\alpha}$ are the Dirac matrices and $\bf {a_{\lambda}}$ is the
$\lambda$ polarization unit vector of the photon vector potential,
with $a_{\pm} = 1/\sqrt{2} (1, \pm i, 0)$,
$a_{\parallel}=(0,0,1)$. Here, $+$ and $-$ denote, respectively, left
and right circular photon polarizations with respect to the
magnetization direction in the solid. Then, x-ray magnetic circular
and linear dichroisms are given by $\mu_{+}-\mu_{-}$ and
$\mu_{\parallel}-(\mu_{+}+\mu_{-})/2$, respectively.  More detailed
expressions of the matrix elements in the electric dipole
approximation may be found in
Refs.~\cite{GET+94,book:AHY04,AHS+04}.  The matrix elements due
to magnetic dipole and electric quadrupole corrections are presented
in Ref.~\cite{AHS+04}.

X-ray absorption spectrscopy (XAS) provides one of the methods to quantify the
strength of the SO interaction in compounds. Van der Laan and Thole showed
that the so-called branching ratio BR = $I_{L_3}/I_{L_2}$ ($I_{L_{2,3}}$ is
the integrated intensity of isotropic XAS at the $L_{2,3}$ edges) is an
important quantity associated with SOC \cite{LaTh88}. The BR is directly
related to the ground-state expectation value of the angular part of the
spin-orbit coupling $<{\bf L \cdot S}>$ through BR = $(2 + r)/(1 - r)$, where
$r$= $<{\bf L \cdot S}>/n_h$ and $n_h$ is the number of holes in 5$d$ states
\cite{LaTh88}. As a result, XAS provides a direct probe of SOC, which is
complementary to other techniques such as the magnetic susceptibility,
electron paramagnetic resonance, and M{\"o}ssbauer spectroscopy (which probe
SOC through the value of the Lande g-factor).

\subsection{RIXS}  

In the direct RIXS process \cite{AVD+11} the incoming photon with energy
$\hbar \omega_{\mathbf{k}}$, momentum $\hbar \mathbf{k}$, and polarization
$\bm{\epsilon}$ excites the solid from ground state $|{\rm g}\rangle$ with
energy $E_{\rm g}$ to intermediate state $|{\rm I}\rangle$ with energy
$E_{\rm I}$. During relaxation an outgoing photon with energy $\hbar
\omega_{\mathbf{k}'}$, momentum $\hbar \mathbf{k}'$ and polarization
$\bm{\epsilon}'$ is emitted, and the solid is in state $|f \rangle$ with
energy $E_{\rm f}$. As a result, an excitation with energy $\hbar \omega =
\hbar \omega_{\mathbf{k}} - \hbar \omega_{\mathbf{k}'}$ and momentum $\hbar
\mathbf{q}$ = $\hbar \mathbf{k} - \hbar \mathbf{k}'$ is created.  Our
implementation of the code for the calculation of the RIXS intensity uses
Dirac four-component basis functions \cite{NKA+83} in the perturbative
approach \cite{ASG97}. RIXS is a second-order process, and its intensity is
given by

\begin{eqnarray}
I(\omega, \mathbf{k}, \mathbf{k}', \bm{\epsilon}, \bm{\epsilon}')
&\propto&\sum_{\rm f}\left| \sum_{\rm I}{\langle{\rm
    f}|\hat{H}'_{\mathbf{k}'\bm{\epsilon}'}|{\rm I}\rangle \langle{\rm
    I}|\hat{H}'_{\mathbf{k}\bm{\epsilon}}|{\rm g}\rangle\over
  E_{\rm g}-E_{\rm I}} \right|^2 \nonumber \\ && \times
\delta(E_{\rm f}-E_{\rm g}-\hbar\omega),
\label{I1}
\end{eqnarray}
where the RIXS perturbation operator in the dipole approximation is given by
the lattice sum $\hat{H}'_{\mathbf{k}\bm{\epsilon}}=
\sum_\mathbf{R}\hat{\bm{\alpha}}\bm{\epsilon} \exp(-{\rm
  i}\mathbf{k}\mathbf{R})$, where $\bm{\alpha}$ are the Dirac matrices. The
sum over intermediate states $|{\rm I}\rangle$ includes the contributions
from different spin-split core states at the given absorption edge. The matrix
elements of the RIXS process in the frame of the fully relativistic Dirac LMTO
method are presented in Ref. \cite{AKB22a}.

\subsection{Crystal structure} 

\begin{figure}[tbp!]
\begin{center}
\includegraphics[width=0.90\columnwidth]{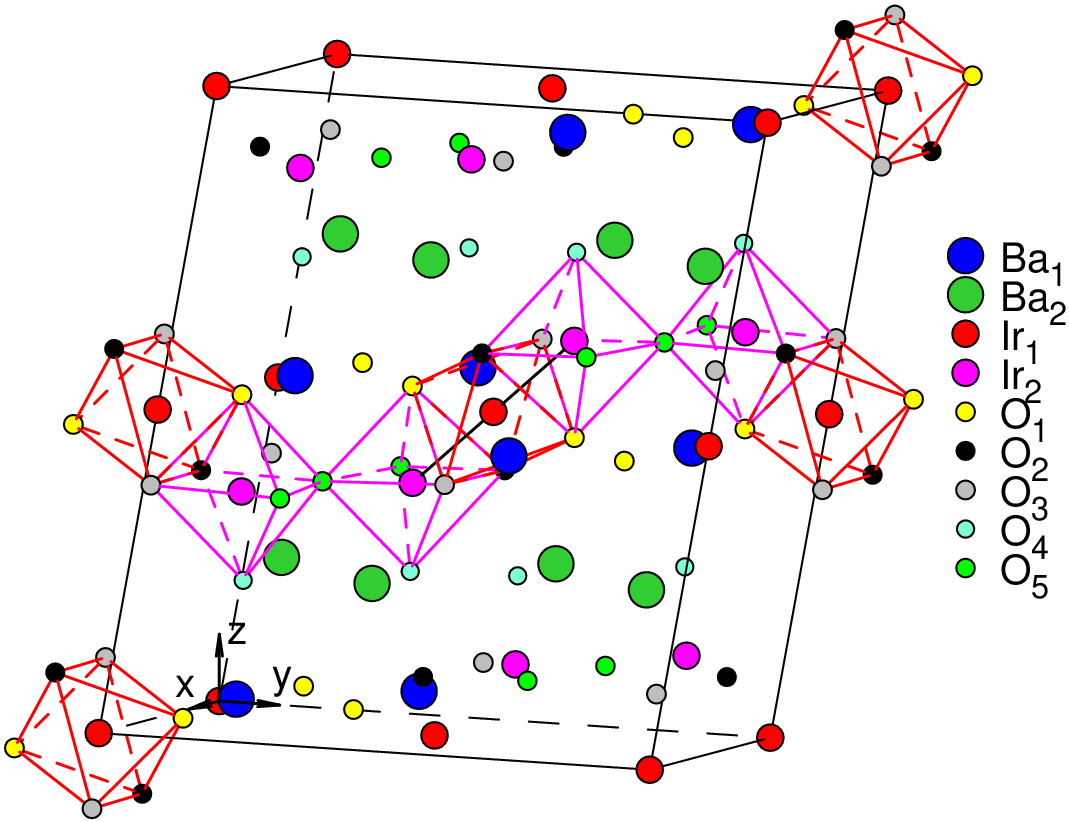}
\end{center}
\caption{\label{struc_BIO}(Color online) The schematic representation of
 Ba$_4$Ir$_3$O$_{10}$ in the monoclinic structure with the space
group $P2_1/a$ (No. 14). }
\end{figure}

Ba$_4$Ir$_3$O$_{10}$ has a monoclinic structure with the space group $P2_1/a$
(No. 14). Its unit cell contains two inequivalent Ba sites, two inequivalent
Ir sites, and five inequivalent O sites (Table \ref{struc_tab_BIO})
\cite{CHW+21}. The structure is built out of Ir$_3$O$_{12}$ trimers, which are
formed by three face sharing IrO$_6$ octahedra (see Fig. \ref{struc_BIO}). The
two Ir atoms, Ir$_1$ and Ir$_2$, occupy the center and outer positions of the
trimer, respectively.  Such trimeric building blocks are connected by corners.
This arrangement leads to the formation of zigzag chains of Ir$_3$O$_{12}$
trimers running along the $b$ axis.

\begin{table}[tbp!]
  \caption {The Wyckoff positions (WP) of Ba$_4$Ir$_3$O$_{10}$ for the
    monoclinic $P2_1/a$ crystal structure (lattice constants $a$ = 5.788 \AA,
    $b$ = 13.222 \AA, $c$ = 7.242 \AA\, and $\beta$ = 112.98$^{\circ}$
    \cite{WM91}). }
\label{struc_tab_BIO}
\begin{center}
\begin{tabular}{|c|c|c|c|c|c|}
\hline
Structure      & WP & Atom     & $x$      & $y$    & $z$     \\
\hline
                   & $4e$   & Ba$_1$    &  0.9695   & 0.6387 & 0.9259 \\
                   & $4e$   & Ba$_2$    &  0.2640   & 0.6131 & 0.5198 \\
                   & $2a$   & Ir$_1$    &  0        & 0      &  0      \\
$P2_1/a$           & $4e$   & Ir$_2$    &  0.1339   & 0.8505 & 0.2496 \\
Ref. \cite{WM91}   & $4e$   & O$_1$     &  0.0179   & 0.1500 & 0.0647   \\
                   & $4e$   & O$_2$     &  0.6665   & 0.0291 & 0.7829  \\
                   & $4e$   & O$_3$     &  0.1244   & 0.0382 & 0.7829  \\
                   & $4e$   & O$_4$     &  0.7278   & 0.1531 & 0.4625  \\
                   & $4e$   & O$_5$     &  0.5861   & 0.2339 & 0.7638  \\
\hline
\end{tabular}
\end{center}
\end{table}

Ir$_1$ and Ir$_2$ belong to intratrimer face-sharing and intertrimer
corner-sharing octahedra, respectively. Two outside octahedra around the
Ir$_2$ ions of each trimer are strongly distorted, chiefly due to the
connection to the octahedra of adjacent trimers. The middle octahedra around
the Ir$_1$ ions are less distorted. The corresponding Ir$_1$-O$_1$,
Ir$_1$-O$_2$, and Ir$_1$-O$_3$ distances for the monoclinic $P2_1/a$ structure
are equal to 2.0312, 1.9941, and 2.0321 \AA, respectively. The Ir$_2$-O$_1$,
Ir$_2$-O$_2$, Ir$_2$-O$_3$, Ir$_2$-O$_4$, and Ir$_2$-O$_5$ distances are equal
to 2.0956, 2.0338, 2.0448, 1.9201, and 1.9777 \AA, respectively. The shortest
Ir-Ir bond is the one between Ir$_1$ and Ir$_2$ within the trimers, which has
a length of 2.5849 \AA, shorter than that of the metallic iridium ($\sim$2.71
\AA \cite{book:Wyck63}).  The average bond lengths are equal to 2.019 \AA\,
and 2.008 \AA\, for the octahedra around Ir$_1$ and Ir$_2$ ions, respectively.

\subsection{Calculation details}

The details of the computational method are described in our previous papers
\cite{AJY+06,AHY+07b,AYJ10,AKB22a} and here we only mention several
aspects. The band structure calculations were performed using the fully
relativistic LMTO method \cite{And75,book:AHY04}. This implementation of the
LMTO method uses four-component basis functions constructed by solving the
Dirac equation inside an atomic sphere \cite{NKA+83}. The exchange-correlation
functional of a GGA-type was used in the version of Perdew, Burke and
Ernzerhof \cite{PBE96}. The Brillouin zone integration was performed using the
improved tetrahedron method \cite{BJA94}. The basis consisted of Ir and Ba
$s$, $p$, $d$, and $f$; and O $s$, $p$, and $d$ LMTO's.

To consider the electron-electron correlation effects we used the
relativistic generalization of the rotationally invariant version of the
GGA+$U$ method \cite{YAF03} which considers that, in the presence of
spin-orbit coupling, the occupation matrix of localized electrons becomes
nondiagonal in spin indexes. Hubbard $U$ was considered an external parameter
and varied from 0.65 to 3.65 eV. We used the value of Hund's coupling 
$J_H$=0.65 eV obtained from constrained LSDA
calculations \cite{DBZ+84,PEE98}. Thus, the parameter $U_{\rm{eff}}=U-J_H$,
which roughly determines the splitting between the lower and upper Hubbard
bands, varied between 0 and 3.0 eV. We adjusted the value of $U$ to achieve
the best agreement with the experiment.

In the RIXS process, an electron is promoted from a core level to an
intermediate state, leaving a core hole. As a result, the electronic structure
of this state differs from that of the ground state. To reproduce the
experimental spectrum, the self-consistent calculations should be carried out
including the core hole. Usually, the core-hole effect has no impact on the
shape of XAS at the $L_{2,3}$ edges of 5$d$ systems and has just a minor
effect on the XMCD spectra at these edges \cite{book:AHY04}. However, the core
hole has a strong effect on the RIXS spectra in transition metal compounds
\cite{AKB22a,AKB22b}, and we consider it in our calculations.

Note, that in our calculations, we rely on the experimentally measured atomic
positions and lattice constants, because they are well established for this
material and still probably more accurate than those obtained from DFT.

\section{Electronic and magnetic structures}
\label{sec:bands}

\begin{figure}[tbp!]
\begin{center}
\includegraphics[width=0.8\columnwidth]{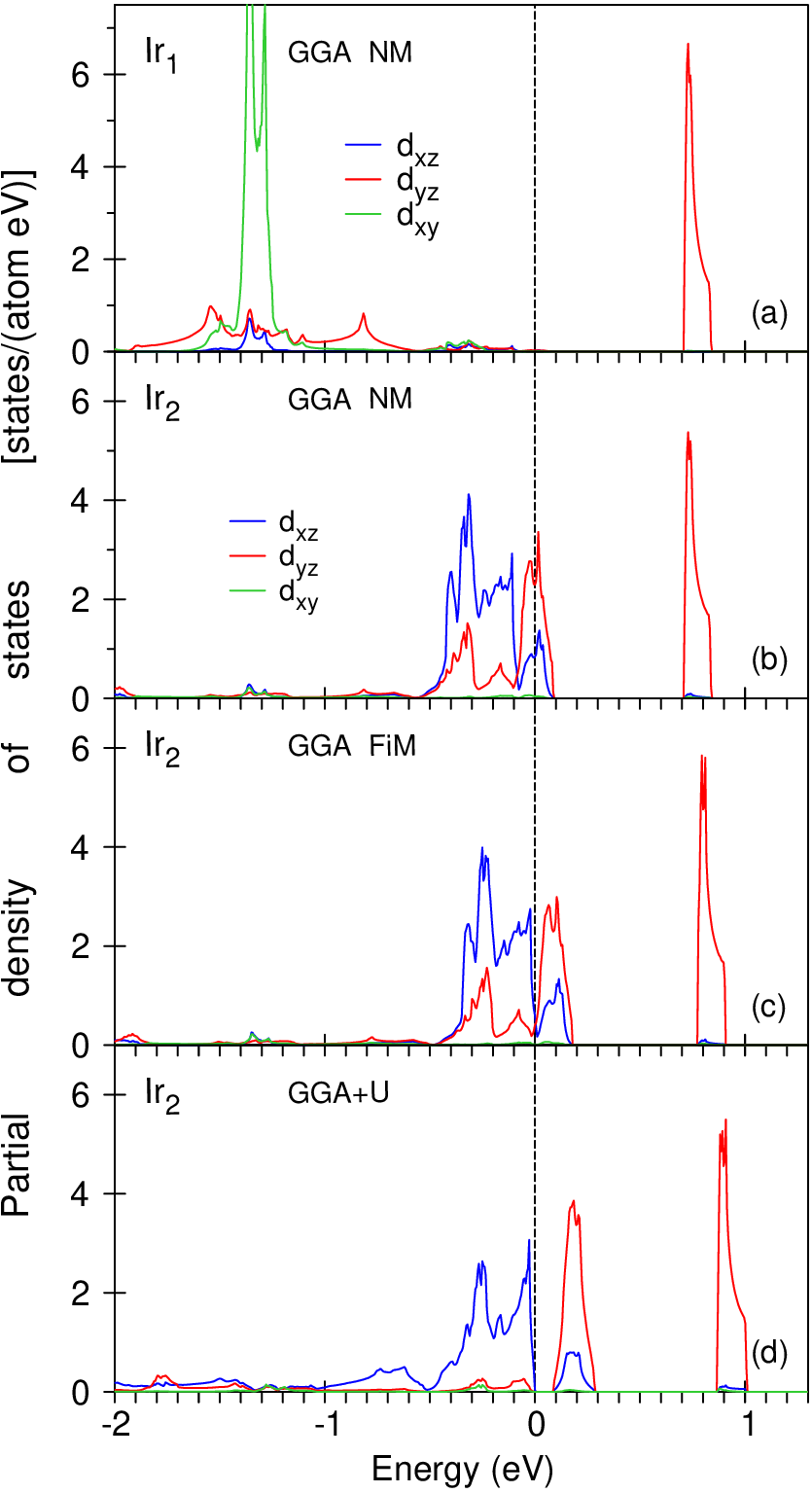}
\end{center}
\caption{\label{Orbitals_BIO}(Color online) The orbital-resolved Ir {\tg}
  partial density of states [in states/(atom eV)] for Ba$_4$Ir$_3$O$_{10}$
  calculated in the GGA for the nomagnetic solution at the Ir$_1$ site (a),
  and the Ir$_2$ site (b). The results of the ferrimagnetic (FiM) solution (c)
  and the GGA+$U$ results (d) for the Ir$_2$ site. }
\end{figure}

We performed GGA, GGA+SO, and GGA+SO+$U$ calculations of the electronic and
magnetic structures of Ba$_4$Ir$_3$O$_{10}$ for the experimental crystal
structure \cite{WM91} (see Table \ref{struc_tab_BIO}). The crystal field at
the Ir site ($Ci$ and $C1$ point symmetry for Ir$_1$ and Ir$_2$ sites,
respectively) causes the splitting of 5$d$ orbitals into five singlets {\dzz},
{\dxxyy}, {\dxy}, {\dxz}, and {\dyz}. In Fig.  \ref{Orbitals_BIO} the
calculated orbital-resolved {\tg} partial density of states (DOS) for
Ba$_4$Ir$_3$O$_{10}$ is presented.  Both the Ir$_1$ and Ir$_2$ ions possess
strong DOS peaks at $\sim$0.8 eV, which indicate tetravalent Ir$^{4+}$ states
(5d$^5$).  The partial DOS is quite different for two nonequivalent Ir sites.

Every Ir$_1$ ion has two neighboring Ir ions in its close vicinity at
$\sim$2.58 \AA\, \cite{CZZ+20,SSL02}, while Ir$_2$ ions have only one
neighboring Ir ion at the same distance (see Fig. \ref{struc_BIO}). As a
results, the Ir$_1$ ions have a tendency of forming quasimolecular orbital
(QMO) states and dimerization.  The {\dxy}, {\dyz}, and {\dxz} orbitals
represent the character of the bands for the {\tg} states at the Ir$_1$
site. The {\dyz} orbital, directed along the trimer bond, provides the
dominant character of the two subbands stemming from {\tg}-derived bands: the
lowest (-1.5 eV) occupied subband and the highest (+0.8 eV) empty subband. The
two subbands with predominant {\dyz} character can be assigned to the bonding
and antibonding states of Ir$_1$ dimer molecules with a splitting energy of
$\sim$2.3 eV. In stark contrast, there are strong peaks at the Fermi level
from the {\dyz} and {\dxz} orbitals at the Ir$_2$ sites in the GGA approach
for the nonmagnetic case [see Fig. \ref{Orbitals_BIO}(b)]. They determine a
metallic solution for the ground state of Ba$_4$Ir$_3$O$_{10}$. The
ferrimagnetic (FiM) solution produces a quasi-gap at the Fermi level [see
  Fig. \ref{Orbitals_BIO}(c)], although the system is still in a metallic
state. However, electrical resistivity shows a clear insulating state across
the entire temperature range measured up to 400 K \cite{CZZ+20}.  For an
insulating solution for the ground state of Ba$_4$Ir$_3$O$_{10}$ one has to
take into account strong Coulomb correlations at the Ir$_2$ site. Due to
relatively small DOS at the Fermi level for the FiM solution, even small
Hubbard $U_{\rm{eff}}$ $\geq$0.35 eV applied only for the Ir$_2$ site is able
to produce an energy gap and an insulating ground state for
Ba$_4$Ir$_3$O$_{10}$ [Fig. \ref{Orbitals_BIO}(d)]. Here we have quite a unique
situation when the insulating state in Ba$_4$Ir$_3$O$_{10}$ is driven by both
the dimerization at the Ir$_1$ site and Mott insulating behavior at the Ir$_2$
one.

The dimerization of transition-metal ions has been frequently seen, for
example, in a wide variety of honeycomb-based 3$d$, 4$d$, and 5$d$ oxides and
halides, including $\alpha$-TiCl$_3$ \cite{Oga60}, $\alpha$-MoCl$_3$
\cite{MYL+17}, Li$_2$RuO$_3$ \cite{MYS+07}, $\alpha$-RuCl$_3$ \cite{AKK+17},
Na$_2$IrO$_3$ \cite{MJF+12,FJM+13}, and the high pressure
$\beta$-Li$_2$IrO$_3$ iridate \cite{AUU18,TKG+19,AKU+21}.

\begin{figure}[tbp!]
\begin{center}
\includegraphics[width=0.99\columnwidth]{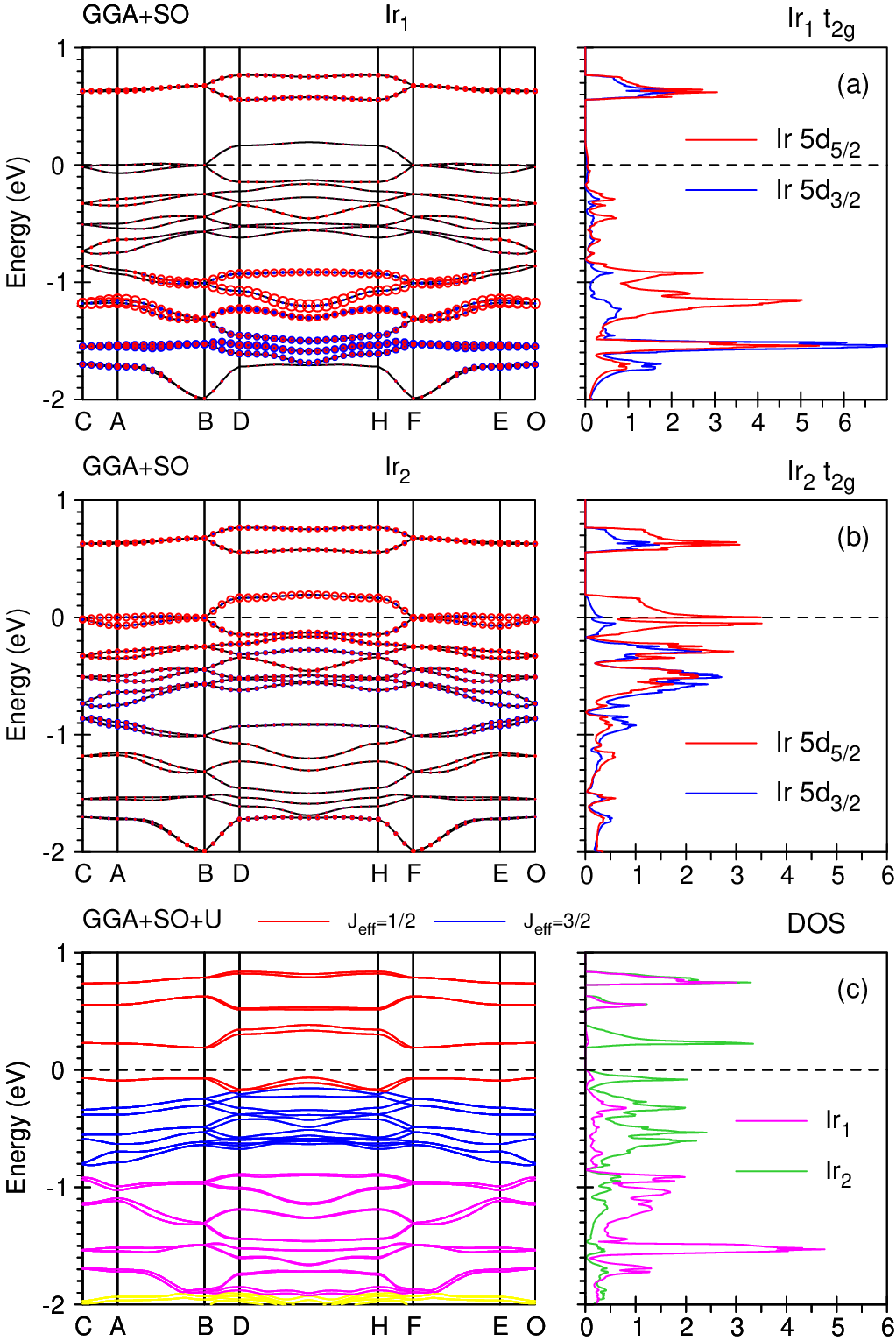}
\end{center}
\caption{\label{BND_Jeff_BIO}(Color online) The energy band structure
  of Ba$_4$Ir$_3$O$_{10}$ and 5$d$ partial density of states for
  Ir$_1$ (the upper panel) and Ir$_2$ (the middle panel) calculated in
  the fully relativistic Dirac GGA+SO approximation. The circles are
  proportional in size to their orbital character projected onto the
  basis set of Ir $d_{3/2}$ (the relativistic quantum number $\kappa$
  = 2, the blue curve) and $d_{5/2}$ ($\kappa$ = $-$3, the red curve)
  states; (the lower panel) the energy bands calculated in the
  GGA+SO+$U$ approximation with $U_{\rm{eff}}$ = 1.3 eV. }
\end{figure}

Figures \ref{BND_Jeff_BIO}(a) and \ref{BND_Jeff_BIO}(b) show the fully
relativistic GGA+SO {\tg} bands and partial DOS for Ir$_1$ and Ir$_2$ sites,
respectively, presented by circles proportional in size to their orbital
character projected onto the basis set of Ir $d_{3/2}$ (the relativistic
quantum number $\kappa$ = 2, the blue curve) and $d_{5/2}$ ($\kappa$ = $-$3,
the red curve) states. According to the partial DOS, the occupied energy bands
for the Ir$_1$ and Ir$_2$ sites are well separated and lie in the energy
intervals from  $-$2 to $-$0.85 eV and from $-$0.85 eV to E$_F$,
respectively. In the strong SOC limit, the {\tg} band at the Ir$_2$ site
splits into effective total angular momentum $J_{\rm{eff}}$ = 1/2 doublet and
$J_{\rm{eff}}$ = 3/2 quartet bands. The functions of the $J_{\rm{eff}}$ = 3/2
quartet are dominated by $d_{3/2}$ states with some weight of $d_{5/2}$
ones. The $J_{\rm{eff}}$ = 1/2 functions are almost completely given by the
linear combinations of $d_{5/2}$ states [Fig. \ref{BND_Jeff_BIO}(a)]. The
GGA+SO+$U$ approach in agreement with the experiment \cite{CZZ+20} gives an
insulating ground state for Ba$_3$InIr$_2$O$_9$ [Fig. \ref{BND_Jeff_BIO}(c)]
with three narrow empty peaks just above the Fermi level. We found that the
best agreement between the calculated and experimentally measured RIXS spectra
at the Ir $L_3$ edge can be achieved for $U_{\rm{eff}}$ = 1.3 eV (see Section
IV).

\begin{figure}[tbp!]
\begin{center}
\includegraphics[width=0.99\columnwidth]{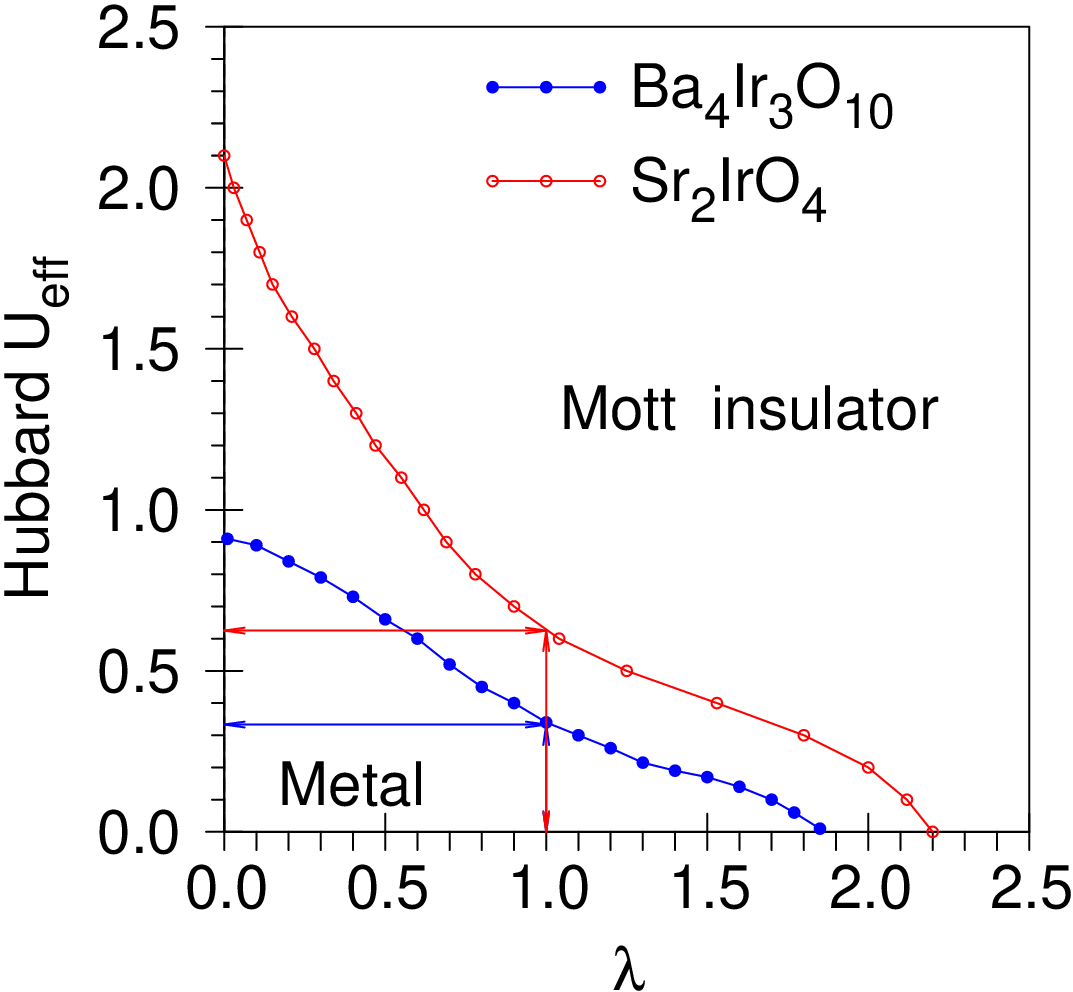}
\end{center}
\caption{\label{U_SO_BIO}(Color online) The phase diagram in the
  $U_{\rm{eff}}-$SOC plane for Ba$_4$Ir$_3$O$_{10}$ (the blue curve) in comparison
  with Sr$_2$IrO$_4$ \cite{AKB24a} (the red curve). The curves separate metal
  (under the curves) and Mott insulator (above the curves) states for
  Ba$_4$Ir$_3$O$_{10}$ and Sr$_2$IrO$_4$, connected via a first-order phase
  transition. The horizontal lines indicate the value of Hubbard $U_{\rm{eff}}$ for
  which the energy band gap opens up. }
\end{figure}

To better understanding the influence of on-site Coulomb interaction and the
spin-orbit coupling in the electronic structure and the energy band formation
in Ba$_4$Ir$_3$O$_{10}$ we present in Figure \ref{U_SO_BIO} a phase diagram in
the $U_{\rm{eff}}-$SOC plane for Ba$_4$Ir$_3$O$_{10}$ in comparison with
Sr$_2$IrO$_4$ \cite{AKB24a}. To obtain this diagram we tuned the SOC term for
the Ir 5$d$ orbitals. A scaling factor $\lambda$ in the SOC term of the
Hamiltonian is introduced in the second variational step \cite{KoHa77}. In
this way, we can enhance the effect of SOC by taking $\lambda$ $>$ 1 or reduce
it by taking $\lambda$ $<$ 1. For $\lambda$ = 0 there is no SOC at all, while
$\lambda$ = 1 refers to the self-consistent reference value. The blue
solid-circled and red solid-circled curves in Fig. \ref{U_SO_BIO} separate
metal (under the curves) and Mott insulator (above the curves) states
connected via a first-order phase transition for Ba$_4$Ir$_3$O$_{10}$ and
Sr$_2$IrO$_4$, respectively. In the case of Sr$_2$IrO$_4$, the energy gap
opens up for $\lambda$ = 0 with $U^c_{\rm{eff}}$ = 2.1 eV and for $\lambda$ =
2.2 with $U^c_{\rm{eff}}$ = 0 eV, the greater the value of $U_{\rm{eff}}$, the
lower the value of $\lambda$ is for the phase transition. For
Ba$_4$Ir$_3$O$_{10}$ the energy gap opens up for $\lambda$ = 0 with
$U^c_{\rm{eff}}$ = 0.91 eV, and for $\lambda$ = 1.85 with $U^c_{\rm{eff}}$ = 0
eV. We can see that Ba$_4$Ir$_3$O$_{10}$ possesses a larger $U_{\rm{eff}}-$SOC
Mott-dielectric phase space in comparison with Sr$_2$IrO$_4$.

\begin{figure}[tbp!]
\begin{center}
\includegraphics[width=0.99\columnwidth]{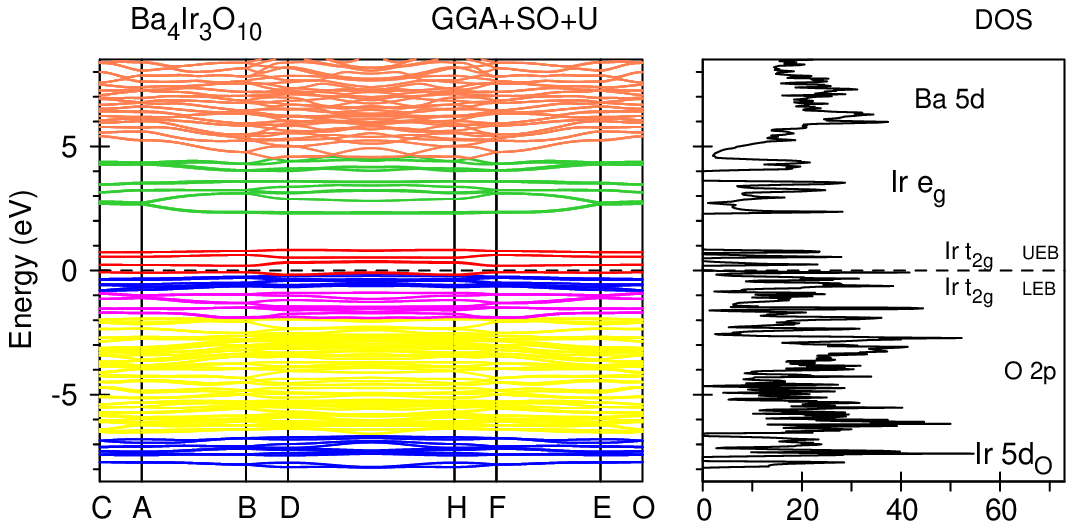}
\end{center}
\caption{\label{BND_BIO}(Color online) The energy band
  structure and total density of states (DOS) [in states/(cell eV)]
  for Ba$_4$Ir$_3$O$_{10}$ calculated in the GGA+SO+$U$ approach
  ($U_{\rm{eff}}$= 1.3 eV). }
\end{figure}

\begin{figure}[tbp!]
\begin{center}
\includegraphics[width=0.99\columnwidth]{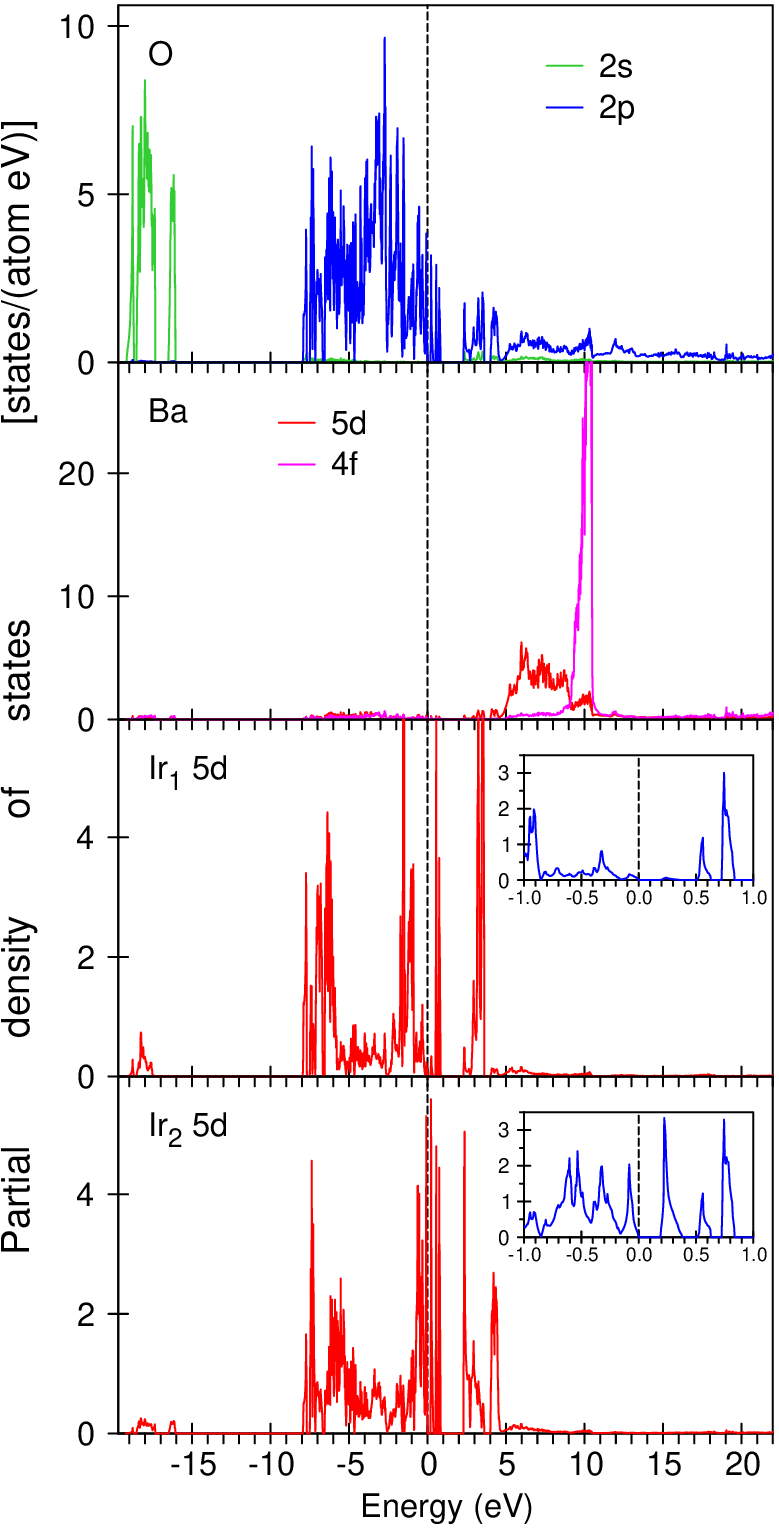}
\end{center}
\caption{\label{PDOS_BIO}(Color online) The partial density of states
  (DOS) [in states/(atom eV)] for Ba$_4$Ir$_3$O$_{10}$ calculated in
  the GGA+SO+$U$ approach ($U_{\rm{eff}}$ = 1.3 eV). }
\end{figure}

Figures \ref{BND_BIO} and \ref{PDOS_BIO} present the energy band structure and
partial DOS, respectively, calculated for Ba$_4$Ir$_3$O$_{10}$ in the
GGA+SO+$U$ approach with $U_{\rm{eff}}$ = 1.3 eV. The occupied {\tg} states
[the low energy band (LEB)] are situated in the energy interval from $-$0.85
eV to $E_F$. The empty {\tg} states [the upper energy band (UEB)] consist of
three narrow single peaks divided by energy gaps and occupy the energy range
from 0.25 to 0.85 eV. There is a significant amount of Ir 5$d$ DOS located at
the bottom of oxygen 2$p$ states from $-$7.9 to $-$4.5 eV below the Fermi
energy. These so called Ir 5$d_{\rm{O}}$ states are provided by the tails of
oxygen 2$p$ states inside the Ir atomic spheres and play an essential role in
the RIXS spectrum at the Ir $L_3$ edge (see Section IV).

The Ba 5$d$ states occupy the energy region from 4.8 to 9.1 eV above the Fermi
energy. A narrow and intensive DOS peak of Ba 4$f$ states is located just
above the Ba 5$d$ states from 9.1 to 10.8 eV. The oxygen 2$s$ states are
situated far below the Fermi level from $-$19.3 to $-$16.1 eV. The occupied O
2$p$ states are localized from $-$7.9 eV to $E_F$. They are strongly
hybridized with Ir 5$d$ states. The empty oxygen 2$p$ states are strongly
hybridized with Ir {\tg} UEB just above the Fermi level and with the Ir {\eg}
states. They are also hybridized with Ba 5$d$ and 4$f$ states.

\begin{table}[tbp!]
  \caption{\label{mom_BIO} The theoretically calculated in the
    GGA+SO+$U$ approach ($U_{\rm{eff}}$ = 1.3 eV) spin $M_s$, orbital
    $M_l$, and total $M_{tot}$ magnetic moments (in {\mb}) in
    Ba$_4$Ir$_3$O$_{10}$ for the FiM solution. }
\begin{center}
\begin{tabular}{ccccccc}
\hline
 atom  & $M_s$ & $M_l$ &  $M_{tot}$ \\
\hline
 Ba$_1$ & -0.0016  &   0.0003  &  -0.0013  \\
 Ba$_2$ & -0.0015  &   0.0014  &  -0.0001  \\
 Ir$_1$ &  0.0590  &   0.0026  &   0.0616 \\
 Ir$_2$ & -0.0748  &  -0.0471  &  -0.1219  \\
 O$_1$  &  0.0013  &  -0.0034  &  -0.0021  \\
 O$_2$  & -0.0026  &   0.0031  &   0.0005  \\
 O$_3$  & -0.0028  &   0.0024  &  -0.0004  \\
 O$_4$  & -0.0235  &  -0.0208  &  -0.0443  \\
 O$_5$  & -0.0103  &  -0.0068  &  -0.0171  \\
\hline
\end{tabular}
\end{center}
\end{table}

Table \ref{mom_BIO} presents the theoretically calculated in the GGA+SO+$U$
approach ($U_{\rm{eff}}$ = 1.3 eV) spin $M_s$, orbital $M_l$, and total
$M_{tot}$ magnetic moments in Ba$_4$Ir$_3$O$_{10}$ for the FiM solution. The
spin and orbital moments at the Ir$_1$ site in Ba$_4$Ir$_3$O$_{10}$ equal to
0.0590 and 0.0026 {\mb}, respectively. They have the same direction since the
Ir 5$d$ states are more than half filled. The spin and orbital moments at the
Ir$_2$ site equal to $-$0.0748 and $-$0.0471 {\mb}, respectively. The ratio
$M_l/M_s$ is equal to 0.04 and 0.63 for the Ir$_1$ and Ir$_2$ sites,
respectively. The corresponding ratio is equal to 1.68 in the SOC driven
$J_{\rm{eff}}$ = $\frac{1}{2}$ iridate Sr$_2$IrO$_4$ \cite{AKB24}. It
indicates a weaker coupling between the local orbital and spin moments in
Ba$_4$Ir$_3$O$_{10}$ in comparison with Sr$_2$IrO$_4$.

\section{XAS, XMCD AND RIXS SPECTRA}

\subsection{I\lowercase{r} $L_{2,3}$ XAS and XMCD spectra}
\label{sec:xmcd}

\begin{figure}[tbp!]
\begin{center}
\includegraphics[width=0.9\columnwidth]{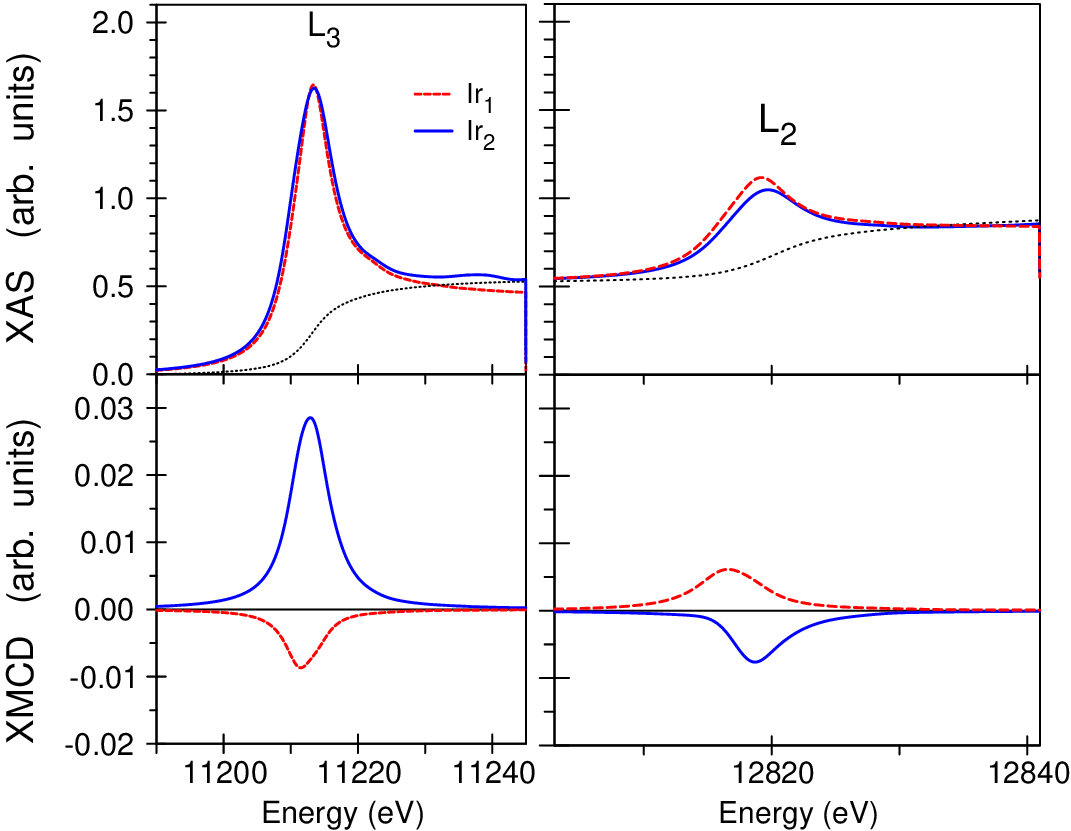}
\end{center}
\caption{\label{xmcd_Ir_BIO}(Color online) The theoretically
  calculated x-ray absorption (the upper panels) and XMCD spectra (the
  lower panels) at the Ir $L_{2,3}$ edges in Ba$_4$Ir$_3$O$_{10}$ for
  the monoclinic $P2_1/a$ crystal structure in the GGA+SO+$U$ approach
  ($U_{\rm{eff}}$ = 1.3 eV). The dotted black curves in the upper
  panels show the background scattering intensity. }
\end{figure}

Figure \ref{xmcd_Ir_BIO} presents the XAS (the upper panel) and XMCD spectra
(the lower panel) at the Ir $L_{2,3}$ edges for Ba$_4$Ir$_3$O$_{10}$
calculated in the GGA+SO+$U$ ($U_{\rm{eff}}$ = 1.3 eV) approach. The isotropic
XAS spectra are dominated by empty $e_g$ states with a smaller contribution
from empty $t_{2g}$ orbitals at lower energy. The XMCD spectra, however,
mainly come from the $t_{2g}$ orbitals. This results in a shift between the
maxima of the XAS and XMCD spectra.

In the limit of negligible SOC effects, the statistical branching ratio BR =
$I_{L_3}/I_{L_2}$ = 2, and the $L_3$ white line is twice the size of the $L_2$
feature \cite{LaTh88}. The theoretically calculated BR in Ba$_4$Ir$_3$O$_{10}$
is 2.45 for the Ir$_1$ site and 3.13 for the Ir$_2$ site. The theoretically
calculated BR for Sr$_2$IrO$_4$ with strong SOC is equal to 3.56
\cite{AKB24a}. Relatively moderate SOC together with molecular orbital-like
states at the Ir$_1$ site in Ba$_4$Ir$_3$O$_{10}$ suggest that the strongly
spin-orbit coupled pure $J_{\rm{eff}}$ = 1/2 picture is not appropriate for
the Ir$_1$ site. It is better use a description based on molecular
orbital-like states. However, the $J_{\rm{eff}}$ = 1/2 model is still valid
for the Ir$_2$ site (see Fig. \ref{BND_Jeff_BIO}).

\subsection{I\lowercase{r} RIXS spectra}
\label{sec:rixs}

\begin{figure}[tbp!]
\begin{center}
\includegraphics[width=0.9\columnwidth]{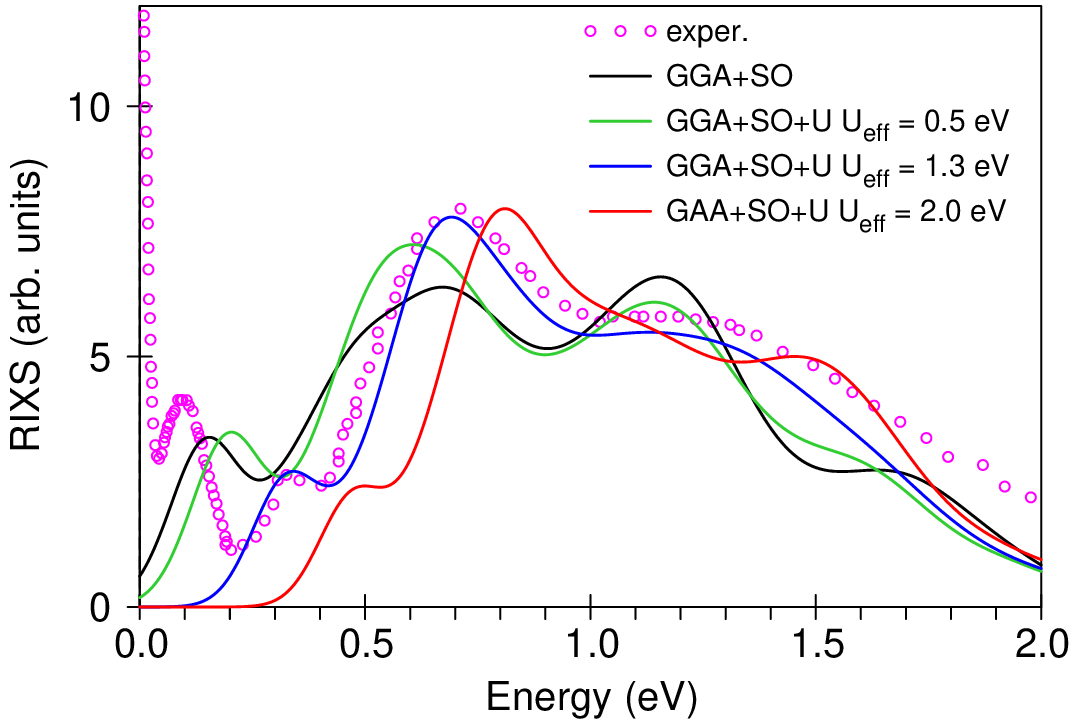}
\end{center}
\caption{\label{rixs_U_BIO}(Color online) The experimental resonance
  inelastic x-ray scattering spectrum (open magenta circles)
  measured by Shen {\it et al.} \cite{SSF+22} at the Ir $L_3$ edge
  in Ba$_4$Ir$_3$O$_{10}$ compared with the theoretically calculated
  ones for different approaches. }
\end{figure}

The experimental RIXS spectrum at the Ir $L_3$ edge was measured by Shen {\it
  et al.} \cite{SSF+22} in the energy range up to 5 eV. In addition to the
elastic peak centered at zero energy loss, the spectrum consists of several
peaks below 2.3 eV and a strong peak at 3.7 eV. We found that the fine
structure situated below 2.3 eV corresponds to intra-{\tg} excitations. These
peaks are very sensitive to the value of the energy gap in
Ba$_4$Ir$_3$O$_{10}$ and the relative position of {\tg} LEB and UEB
(Fig. \ref{BND_BIO}). Figure \ref{rixs_U_BIO} shows the experimental RIXS
spectrum measured by Shen {\it et al.} \cite{SSF+22} compared with the
theoretical spectra calculated for $\tg \to \tg$ transitions in the GGA+SO and
GGA+SO+$U$ approaches for the FiM solution for different $U_{\rm{eff}}$
values. The best agreement was found for $U_{\rm{eff}}$ = 1.3 eV. The GGA+SO
calculations as well as the GGA+SO+$U$ approach with smaller $U_{\rm{eff}}$ do
not produce adequate agreement with the experimental data.  The larger values
of $U_{\rm{eff}}$ shift the RIXS spectrum towards higher energies.

There is a low energy peak at $\sim$0.1 eV which is not produced by our
first-principle RIXS calculations. Shen {\it et al.} \cite{SSF+22} show that
this peak belongs to low-energy magnetic excitations. It is interesting to
note that the reference iridate Sr$_2$IrO$_4$ also possesses a low energy peak
at 0.1 eV, which is considered in Ref. \cite{KDS+14} as magnon excitations in
agreement with scanning tunneling microscope measurements \cite{NBA+14}. The
theoretical description of magnon and exciton spectra demands a many-body
approach beyond the one-particle approximation, such as the Bethe-Salpiter
equation for exciton spectra and calculations of the magnon dispersion and the
electron-magnon interaction for magnon spectra.

\begin{figure}[tbp!]
\begin{center}
\includegraphics[width=0.85\columnwidth]{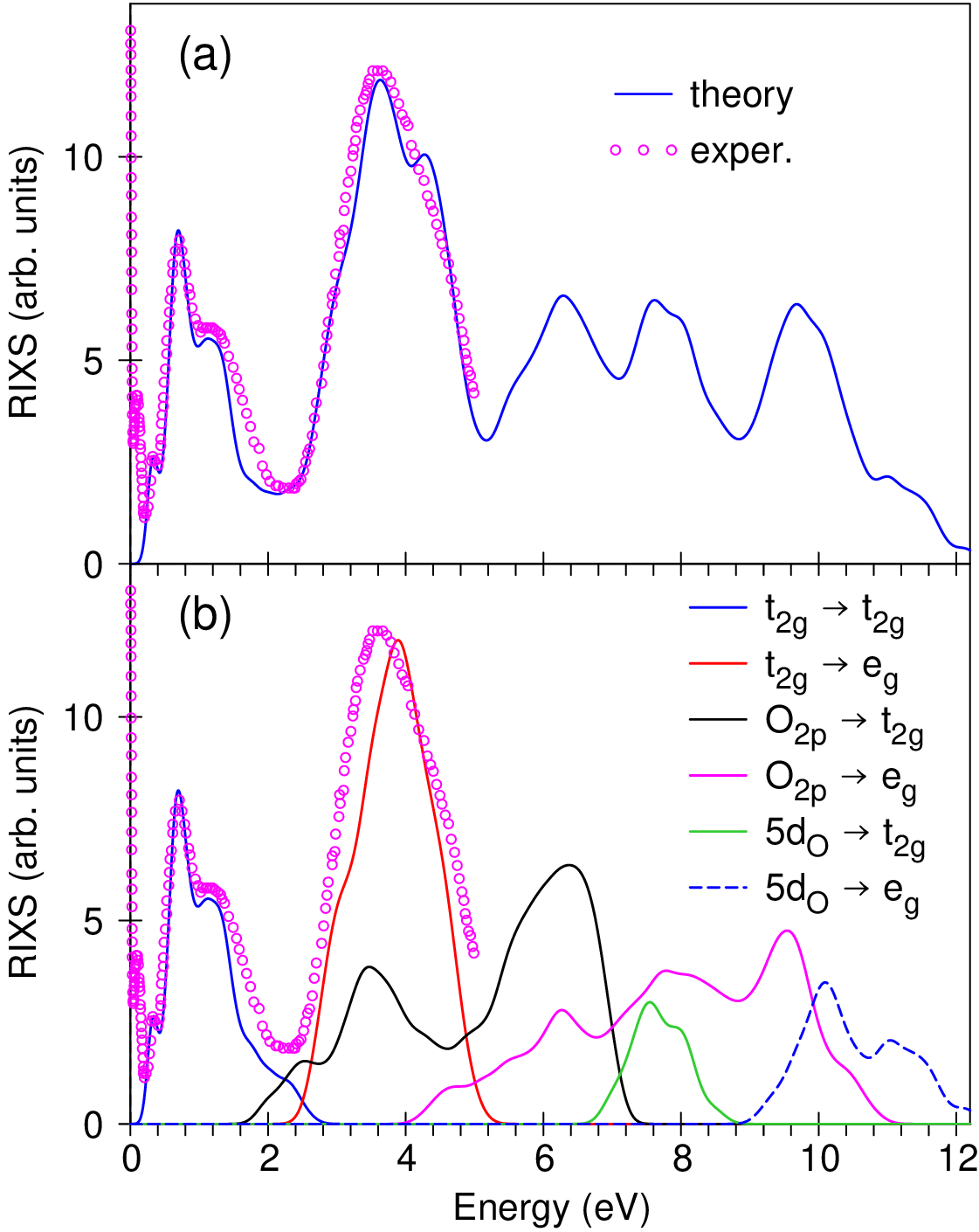}
\end{center}
\caption{\label{rixs_Ir_L3_BIO}(Color online) (a) The experimental
  resonance inelastic x-ray scattering spectrum (open magenta
  circles) measured by Shen {\it et al.} \cite{SSF+22} at the Ir $L_3$
  edge in Ba$_4$Ir$_3$O$_{10}$ compared with the theoretically
  calculated one in the GGA+SO+$U$ approach ($U_{\rm{eff}}$ = 1.3 eV);
  (b) partial contributions from different transitions between the
  energy bands presented in Fig. \ref{BND_BIO} in comparison with the
  experimental spectrum \cite{SSF+22}. }
\end{figure}

Figure \ref{rixs_Ir_L3_BIO}(a) presents the theoretically calculated and
experimentally measured RIXS spectra at the Ir $L_3$ edge for
Ba$_4$Ir$_3$O$_{10}$ \cite{SSF+22} in a wide energy interval up to 12
eV. Figure \ref{rixs_Ir_L3_BIO}(b) shows the partial contributions from
different transitions between the energy bands presented in Fig. \ref{BND_BIO}
in comparison with the experimental spectrum \cite{SSF+22}. As we mentioned
above, the peaks situated below 2.3 eV correspond to intra-{\tg} excitations
(the blue curve in Fig. \ref{rixs_Ir_L3_BIO}). The peak located at $\sim$3.7
eV was found to be due to $\tg \rightarrow \eg$ transitions (the red
curve). The O$_{2p}$ $\rightarrow$ {\tg} transitions (the black curve) also
contribute to this peak. There are three peaks above 5 eV which can be
associated with O$_{2p}$ $\rightarrow$ {\tg} (the black curve) and O$_{2p}$
$\rightarrow$ {\eg} (the magenta curve) transitions, as well as
charge-transfer excitations 5$d_{\rm{O}}$ $\rightarrow \tg$ (the green curve)
and 5$d_{\rm{O}}$ $\rightarrow \eg$ (the blue dashed curve). The theoretical
calculations are in good agreement with the experimental data.

\begin{figure}[tbp!]
\begin{center}
\includegraphics[width=0.9\columnwidth]{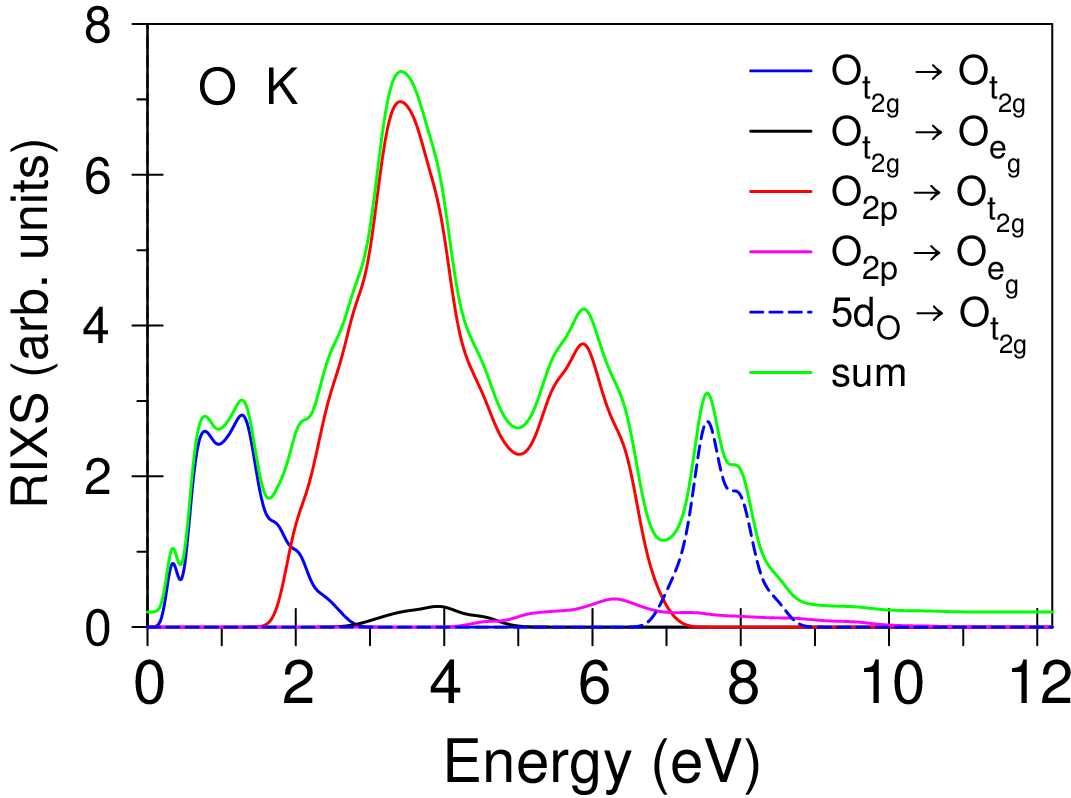}
\end{center}
\caption{\label{rixs_O_K_BIO}(Color online) Partial contributions
  from different transitions between the energy bands at the O $K$
  edge in Ba$_4$Ir$_3$O$_{10}$ calculated in the GGA+SO+$U$ approach
  ($U_{\rm{eff}}$ = 1.3 eV). }
\end{figure}

Figure \ref{rixs_O_K_BIO} shows the theoretically calculated in the GGA+SO+$U$
approach RIXS spectrum at the O $K$ edge in Ba$_4$Ir$_3$O$_{10}$. The fine
structures at $\le$2.3 eV are derived from the interband transitions between
the oxygen 2$p$ states strongly hybridized with the Ir {\tg} states in close
vicinity to the Fermi level: the $O_{\tg} \rightarrow O_{\tg}$ transitions
(the blue curve). However, the $O_{\tg} \rightarrow O_{\eg}$ transitions are
very weak (the black curve). The major two-peak fine structure situated at
2$-$7 eV is quite intensive and derived from the O$_{2p}$ $\rightarrow
O_{\tg}$ transitions (the red curve). The O$_{2p}$ $\rightarrow O_{\eg}$
transitions are less intensive (the magenta curve). The narrow peak at 7$-$9
eV is due to 5$d_{\rm{O}}$ $\rightarrow O_{\tg}$ transitions (the blue dashed
curve). Experimental measurements of the RIXS spectrum at the O $K$ edge in
Ba$_4$Ir$_3$O$_{10}$ are highly desirable.

\section{Conclusions}

To summarize, we have investigated the electronic structure of
Ba$_4$Ir$_3$O$_{10}$ in the frame of the fully relativistic spin-polarized
Dirac approach. We have also presented comprehensive theoretical calculations
of the RIXS spectra at the Ir $L_3$ and oxygen $K$ edges, as well as the XAS
and XMCD spectra at the Ir $L_{2,3}$ edges.

The delicate interplay between electronic correlations, SOC, intersite
hopping, and a crystal field splitting leads to a strongly competing ground
state of Ba$_4$Ir$_3$O$_{10}$. We found that the ground magnetic state of
Ba$_4$Ir$_3$O$_{10}$ is ferrimagnetic. The GGA and GGA+SO approaches give a
metallic ground state in Ba$_4$Ir$_3$O$_{10}$. It is in contradiction with
electric conductivity and energy loss measurements, which firmly indicate an
insulating character of Ba$_4$Ir$_3$O$_{10}$. A quasi-2D structure of
Ba$_4$Ir$_3$O$_{10}$ is composed of buckled sheets, which constitute
corner-connected Ir$_3$O$_{12}$ trimers containing three distorted
face-sharing IrO$_6$ octahedra. The Ir atoms are distributed over two
symmetrically inequivalent sites: the center of the trimer (Ir$_1$) and its
two tips (Ir$_2$). The Ir$_1$ $-$ Ir$_2$ distance within the trimer is quite
small and equals to 2.58\AA\, at low temperature. As a result, the clear
formation of bonding and antibonding states at the Ir$_1$ site occurs. The
large bonding-antibonding splitting stabilizes the $d_{yz}$-orbital-dominant
antibonding state of $t_{2g}$ holes and produces a wide energy gap at the
Fermi level. However, the energy gap opens up in Ba$_4$Ir$_3$O$_{10}$ only
with taking into account strong Coulomb correlations at the Ir$_2$
site. Therefore, we have quite a unique situation when the insulating state in
Ba$_4$Ir$_3$O$_{10}$ is driven by both the dimerization at the Ir$_1$ site and
Mott insulating behavior at the Ir$_2$ one.

The theoretically calculated Ir $L_3$ RIXS spectrum is in good agreement with
the experiment. We found that the low energy part of the RIXS spectrum
$\le$2.3 eV corresponds to intra-{\tg} excitations. The peak located at
$\sim$3.7 eV was found to be due to $\tg \rightarrow \eg$ transitions. The
O$_{2p}$ $\rightarrow$ {\tg} transitions also contribute to this peak There
are three peaks above 5 eV which can be associated with O$_{2p}$ $\rightarrow$
{\tg} and O$_{2p}$ $\rightarrow$ {\eg} transitions, as well as charge-transfer
excitations 5$d_{\rm{O}}$ $\rightarrow \tg$ and 5$d_{\rm{O}}$ $\rightarrow
\eg$.

The major contribution into the RIXS spectrum at the O $K$ edge between 2 and
7 eV comes from the O$_{2p}$ $\rightarrow O_{\tg}$ transitions. The O$_{2p}$
$\rightarrow O_{\eg}$ transitions are less intensive. We found that the
$O_{\tg} \rightarrow O_{\tg}$ transitions are situated below 2.3 eV, and the
$O_{\tg} \rightarrow O_{\eg}$ transitions are very weak.  The narrow peak at
7$-$9 eV is due to 5$d_{\rm{O}}$ $\rightarrow O_{\tg}$ transitions.

\section*{Acknowledgments}

The studies were supported by the National Academy of Sciences of Ukraine
within the budget program KPKBK 6541230 "Support for the development of
priority areas of scientific research".
 

\newcommand{\noopsort}[1]{} \newcommand{\printfirst}[2]{#1}
  \newcommand{\singleletter}[1]{#1} \newcommand{\switchargs}[2]{#2#1}

\end{document}